\documentclass[12pt]{article}
\usepackage[cp1251]{inputenc}
\usepackage[russian]{babel}
\usepackage{amsmath}
\usepackage{amssymb}
\usepackage{amsfonts}
\usepackage{cite}
\usepackage{cmap}
\usepackage[pdftex,unicode]{hyperref}

\newcommand{\itemaa}{\smallskip${{\bigstar}}$\hspace{0.1cm}}
\newcounter{inum}
\newcommand{\itemnum}{\addtocounter{inum}{1}\textit{\theinum.\ }} 

\newtheorem{theorem}{Теорема}

\numberwithin{equation}{section}

\newcounter{meq}
\def\mmeq#1{\refstepcounter{meq}\begin{equation*} #1\eqno{(\hbox{S}\themeq)}\end{equation*}}
\def\sref#1{S\ref{#1}}
\def\seqref#1{(S\ref{#1})}

\def\eqa#1{\begin{equation}\begin{aligned}#1\end{aligned}\end{equation}}
\def\eq#1{\begin{equation}#1\end{equation}}
\def\eqs#1{\begin{eqnarray}#1\end{eqnarray}}

\def\seq#1{\begin{equation*}#1\end{equation*}}

\let\bb\mathbb

\def\phi{\varphi}

\let\t\tilde

\newcommand{\arcsinh}{\text{\rm arcsinh}\,}
\newcommand{\arccosh}{\text{\rm arccosh}\,}

\def\be{\begin{equation}}
\def\ee{\end{equation}}
\def\ba{\begin{aligned}} 
\def\ea{\end{aligned}}

\def\R{\bb R}  
  
\def\Q{\bb Q}  

\newcounter{theo}

\newcounter{lem}

\newcounter{prop}
\newcounter{rem}

\newcounter{defi}

\newcounter{examp}

\def\pfrac#1#2{\frac{\partial #1}{\partial #2}}

\numberwithin{equation}{section}
\title{Классификация полудискретных уравнений гиперболического типа. Случай симметрий третьего порядка.}

\author{Р.Н. Гарифуллин.\thanks{\rm Исследование выполнено за счет гранта Российского научного фонда № 21-11-
00006, https://rscf.ru/project/21-11-00006/.}
\thanks{\rm Институт математики с вычислительным центром Уфимского федерального исследовательского центра РАН, Уфа, Россия.}}
\begin{document}

\maketitle {\small
\begin{quote}
\noindent{\bf Аннотация. } В работе проводится квалификация полудискретных уравнений гиперболического типа. Исследуется класс уравнений вида
\seq{\frac{du_{n+1}}{dx}=f\left(\frac{du_{n}}{dx},u_{n+1},u_{n}\right),} здесь неизвестная функция $u_n(x)$ зависит от одной дискретной $n$ и одной непрерывной $x$ переменных. 
Квалификации основывается на требовании существования высших симметрий в дискретном и непрерывной направлениях. Рассматривается случай когда симметрии имеют порядок 3 в обоих направлениях. В результате получен список уравнений с требуемыми условиями.
\medskip

\noindent{\bf Ключевые слова:}  {интегрируемость, высшая симметрия, классификация, полудискретное уравнение, гиперболический тип.
}
\medskip
\end{quote}
 }

В этой работе исследуются  полудискретные уравнения гиперболического типа\eq{u_{n+1,x}=f(u_{n,x},u_{n+1},  u_{n},x )\label{seq},} где неизвестная функция $u_n(x)$ зависит от одной дискретной $n$ и одной непрерывной $x$ переменных. Здесь и ниже используется обозначение $$u_{k,x}=\frac{du_{k}}{dx},\ u_{k,xx}=\frac{d^2u_{k}}{dx^2},\ u_{k,xxx}=\frac{d^3u_{k}}{dx^3}.$$

Наиболее известным представителем этого класса является одевающая цепочка, подробное исследование которой проведено в статье А.П.~Веселова и А.Б.~Шабата  \cite{vs93}\eq{u_{n+1,x}+u_{n,x}=u_{n+1}^2-u_{n}^2\label{dr},} которая возникла как преобразование Бэклунда для модифицированного уравнения Кортевега де Вриза:
\eq{u_{n,t}=u_{n,xxx}-6u_n^2u_{n,x}\label{kdv}.} С другой стороны уравнение \eqref{kdv} можно рассматривать как высшую симметрию уравнения \eqref{dr}. По дискретному направлению высшая симметрия уравнения \eqref{dr} имеет вид \eq{u_{n,\tau}=\frac{1}{u_{n+1}+u_{n}}-\frac{1}{u_{n}+u_{n-1}}\label{vol}} и является известным дифференциально-разностным уравнением \cite{y83,y06}. В статье Р.И.~Ямилова  \cite{y90} был приведен ряд примеров троек уравнений типа \eqref{dr}-\eqref{vol}.

В недавней работе \cite{gh21} был предложен метод построения высших симметрий уравнений вида \eqref{seq}. Было показано, что высшая симметрия в непрерывном направлении является эволюционным уравнением вида:
\eq{u_t=\frac{d^Nu_n}{dt^N}+F\left(x,u_n, \, \frac{du_n}{dt}, \dots, \frac{d^{N-1}u_n}{dt^{N-1}}\right).      \label{scalar}
}Такие уравнения называются уравнениями с постоянной сепарантой \cite{ms12}. С другой стороны полудискретного уравнения вида \eqref{seq} совместностное с уравнением вида \eqref{scalar} можно рассматривать как автопреобреобразование Бэклунда. Поэтому уравнения \eqref{scalar} сами по себе также являются инегрируемыми уравнениями. Список таких уравнений порядков 3 и 5 был приведен в обзоре \cite{ms12}, в нем также подробно изложена история вопроса. Уравнения вида \eqref{scalar} при $N=3$ были проклассифицированы в работе \cite{ss82}.

В этой работе мы будет рассматривать уравнения $S$-интегрируемые{\footnote {Терминология принадлежит F. Calogero}} уравнения третьего порядка и для них искать полудискретные уравнения вида \eqref{seq}. Список таких уравнений третьего порядка имеет вид:
\begin{align}
&u_t=u_{xxx}-3uu_x,\ \label{list1}\\
&u_t=u_{xxx}-6u^2u_x, \label{list2}\\
&u_t=u_{xxx}+6u_x^2, \label{list3}\\
&u_t=u_{xxx}-\frac{1}{2}u_x^3-\frac23(b_1e^{2u}+b_2e^{-2u})u_x, \label{list4}\\
&u_t=u_{xxx}-\frac{3u_xu_{xx}^2}{2(u_x^2+1)}+b_1(u_x^2+1)^{3/2}+b_2u_x^3, \label{list5}\\
&u_t=u_{xxx}-\frac{3}{2}\,\frac{u^{2}_{xx}}{u_{x}}+\frac{Q(u)}{u_{x}}, \label{list7}\\
&u_{t}= u_{xxx}- \frac{3}{8}\frac{\big((Q(u)+u_x^2)_{x}\big)^2}{u_x\,(Q(u)+u_x^2)}+\frac{1}{2} Q''(u)\,u_x, \label{list6}\\
&u_t=u_{xxx}-\frac{3u_{xx}^2}{2u_x}, \label{list8}\\
&u_t=u_{xxx}-\frac{3u_{xx}^2}{4u_x}+b_1u_x^{3/2}-3b_2^2u_x^2,\ \ b_1\ne0\ \mbox{\rm или } b_2\ne0, \label{list9}
\end{align}
Здесь $ Q=b_{0}+b_{1} u+b_{2} u^{2}+b_{3} u^{3}+b_{4} u^{4}$ --- произвольный многочлен, $b_i$ --- произвольные постоянные.

Приведенный список уравнений отличается от списка работы \cite{ms12} переобозначением констант и уравнениями \eqref{list7} и \eqref{list6}, которые приведены в точечно эквивалентном виде без использования функции Вeйерштрасса.

Известно, что при дробно-линейных преобразованиях 
\begin{equation}\label{meb}
u=\frac{z_1 \t u+z_2}{z_3 \t u+z_4}
\end{equation}
многочлен $Q$ меняется по закону 
$$
\t Q(\t u)=Q\left(\frac{z_1 \t u+z_2}{z_3 \t u+z_4}\right) (z_3 \t u+z_4)^4 (z_1 z_4-z_2 z_3)^{-2}.
$$
В зависимости от структуры кратных корней многочлен $Q$ может быть приведен преобразованием (\ref{meb}) и растяжениями $x$ и $t$ 
к одной из следующих канонических форм: $Q(x)=x(x-1)(x-k)$, $Q(x)=x(x-1)$, $Q(x)=x^2$, $Q(x)=x$, $Q(x)=1$ и $Q(x)=0$.

Использование $C$-интегриуемых уравнений приводит только к Дарбу интегрируемым или линейным уравнениям гиперболического типа, которые представляют меньший интерес. 

Высшие симметрии в дискретном направлении являются уравнениями типа Вольтерры:
\eq{u_{n,\tau}=G(u_{n-1},u_{n},u_{n+1}).\label{d_sym}} Полный список интегрируемых уравнений такого типа получен в работе \cite{y83}, более подробное изложение в обзоре \cite{y06}.

В работе ищутся только автономные по дискретной переменной $n$ уравнения вида \eqref{seq} и соответственно используются только автомномные высшие симметрии (\ref{list1}--\ref{list9}). Это, с одной стороны, является упрощением задачи, но, с другой стороны, не известно неавтономных уравнений вида \eqref{seq} совместных с дискретной высшей симметрией (возможно неавтономной) вида \eqref{d_sym}. Отметим, что полученные ответы содержат дополнительные произвольные константы, вместо этих констант можно брать функции зависящие от $n$, при этом останется совместность с непрерывными высшими симметриями. 

В полностью диcкретном случае, сущуствуют автономные уравнения, у которых одна из высший симметрий имеет порядок 3 и является неавтомной, а высшая симметрия в другом направление имеет больший порядок, см. \cite{gy12,gmy14,gy19}. Подобные уравнения в этой работе не исследуются. Вопрос об их существование остается открытым.

\section{Метод исследования.}

Из требования совместности уравнений \eqref{seq} и \eqref{scalar} получаем определяющее уравнение
\eq{V_{n+1,x}=\pfrac{f}{u_{n,x}}V_{n,x}+\pfrac{f}{u_{n+1}}V_{n+1}+\pfrac{f}{u_{n}}V_n, \label{lineq}}
где через $V_n$ обозначена правая часть уравнения \eqref{scalar}. Здесь используются обозначения:
\seq{V_{n,x}=\frac{d}{dx}V_n,\quad V_{n+1,x}=\frac{d}{dx}V_{n+1}}

Если функция $f$, определяющая правую часть уравнения \eqref{seq} известна, то из уравнения \eqref{lineq} можно находить правую часть высшей симметрии \eqref{scalar}. Эта процедура подробно описана в работе \cite{gh21}. Здесь же наоборот известна высшая симметрия, а само полудискретное уравнение не задано. Поэтому на функцию $f$ получается сложное нелинейное уравнение. Однако уравнение содержит дополнительные переменные, от которых не зависит функция $f$. Наличие этих переменных позволяет получать более простые дифференциальные следствия и определить неизвестную функцию $f$. 

\section{Результаты классификации.}
В данной секции приводятся найденные полудискретные уравнения и их высшие симметрии в дискретном направлении. Они сгруппированы по виду высшей симметрии в $x$ направлении. Верно следующее утверждение:
\begin{theorem} Если невырожденное нелинейное автономное уравнение \eqref{seq} допускает непрерывную высшую симметрию в виде одного из уравнений (\ref{list1}--\ref{list9}), то оно имеет вид \seqref{hkdv} -- \seqref{hyp9}.
Все уравнения списка \seqref{hkdv} -- \seqref{hyp9} обладают дискретными высшими симметрями вида \eqref{d_sym}.
\end{theorem}

{\bf Схема доказательства. } Правые части уравнений (\ref{list1}--\ref{list9}) брались в качестве функции $V_n$ в определеяющем уравнении \eqref{lineq}. Для каждого из этих уравнений находились все возможные функции $f$ -- правые части уравнений \eqref{seq}. Для полученных уравнений вида \eqref{seq} находились дискретные симметрии. Ниже приводится список найденных полудискретных уравнений (для них используется спициальная нумерация вида (S...), и их дискетные высшие симметрии. Разные уравнения списка (\ref{list1}--\ref{list9}) разделяются с использованием символа ${{\bigstar}}$.

\itemaa 
Для уравнения \eqref{list1} полудискретное уравнение имеет вид: 
\mmeq{{u_{n+1,x}+u_{n,x}=\sqrt{u_{n+1}+u_{n}+2a}(u_{n+1}-u_{n})\label{hkdv}}}
\eq{u_{n,\tau}=\frac{\sqrt{u_{n+1}+u_{n}+2a}-\sqrt{u_{n}+u_{n-1}+2a}}{\sqrt{u_{n+1}+u_{n}+2a}+\sqrt{u_{n}+u_{n-1}+2a}}\label{sym_tau_1}}
Уравнение \eqref{sym_tau_1} это уравнение (V9) при $P(y^2)=1$ списка из \cite[p.R596]{y06}.

\itemaa Для уравнения \eqref{list2} уравнение вида \eqref{seq} и симметрия в дискретном направлении записываются имеют из следующих видов:  
\mmeq{{u_{n+1,x}+u_{n,x}=\sqrt{(u_{n+1}+u_{n})^2+2a}(u_{n+1}-u_{n})\label{hyp_2}}}
\eqa{u_{n,\tau}=\frac{R(u_{n+1},u_{n},u_{n-1})-R(u_{n+1},u_{n},u_{n+1})^{1/2}R(u_{n-1},u_{n},u_{n-1})^{1/2}}{a(u_{n+1}-u_{n-1})},\\ R(u,v,w)=(u+v)(v+w)+2a,\label{sym_tau_2}}
Уравнение \eqref{sym_tau_2} это (V4) при $\nu=-1$.

\mmeq{{u_{n+1,x}-u_{n,x}=\sqrt{(u_{n+1}-u_{n})^2+2a}(u_{n+1}+u_{n})\label{hyp_2.1}}}
\eqa{u_{n,\tau}=\frac{R(u_{n+1},u_{n},u_{n-1})-R(u_{n+1},u_{n},u_{n+1})^{1/2}R(u_{n-1},u_{n},u_{n-1})^{1/2}}{u_{n+1}-u_{n-1}},\\ R(u,v,w)=(u-v)(w-v)+2a,\label{sym_tau_2.1}}
Уравнения \seqref{hyp_2} и \seqref{hyp_2.1}, а также уравнения \eqref{sym_tau_2} и \eqref{sym_tau_2.1} связаны неавтономной точечной заменой $\tilde u_n=(-1)^nu_n$. 

{\bf{Замечение.}} Одевающая цепочка \eqref{dr} и ее высшая симметрия \eqref{vol} получаются из \seqref{hyp_2} и \eqref{sym_tau_2} в пределе $a\to 0$. Подобный предел в уравнении \seqref{hyp_2.1} приводит к уравнению интегрируемому по Дарбу
$$u_{n+1,x}-u_{n,x}=u_{n+1}^2-u_{n}^2. $$

\itemaa Для уравнения \eqref{list3} получаем: 
\mmeq{{u_{n+1,x}+u_{n,x}=-(u_{n+1}-u_{n})^2+a(u_{n+1}-u_{n})+a_1\label{hyp_3}}}
\eqa{u_{n,\tau}=\frac{1}{u_{n+1}-u_{n-1}+a}+c,\label{sym_tau_3}}
Уравнение \eqref{sym_tau_3} это (V4) при $\nu=0,R(u,v,w)=1$, $c$ отвечает за точечную симметрию, константу $a$ можно сделать равной нулю за счет неавтономного преобразования $\tilde u_{n}=u_n+an/2$.

\itemaa Для уравнения \eqref{list4} имеются 2 случая в зависимости от параметров:

1.  В слаучае $b_1 \neq 0$ или $b_2 \neq 0$ получаем две пары уравнений: 
\mmeq{{u_{n+1,x}=-u_{n,x}+(e^{-u_{n+1}}\pm e^{-u_{n}})\sqrt{b_1e^{2u_{n+1}+2u_{n}}\pm 2ae^{u_{n+1}+u_n}+b_2},\label{hyp_4_1}}}
\eqa{u_{n,\tau}=e^{-u_n}\frac{R(e^{u_{n+1}},e^{u_n},e^{u_{n-1}})-R_1(e^{u_{n+1}},e^{u_n})^{1/2}R_1(e^{u_{n-1}},e^{u_n})^{1/2}}{e^{u_{n+1}}-e^{u_{n-1}}},\\ R(u,v,w)=b_1v^2uw\pm av(u+w)+b_2,\quad R_1(v,u)=R(v,u,w).}
Уравнения с разными знаками связаны заменой $\tilde{u}_{n}=u_{n}+In\pi,\ \tilde x =(-1)^{n+1}x$ (здесь и ниже $I$ -- мнимая единица, т.е. $I^2=-1$),  которая в уравнение \eqref{list4} меняет $\tilde t=(-1)^{n+1}t$.
 \mmeq{{u_{n+1,x}=u_{n,x}+(b_1^{1/2}\pm b_2^{1/2}e^{-u_{n+1}-u_{n}})\sqrt{e^{2u_{n+1}}\pm 2ae^{u_{n+1}+u_{n}}+e^{2u_n}},\label{hyp_4_3}}}
\eqa{u_{n,\tau}=e^{-u_n}\frac{R(e^{u_{n+1}},e^{u_n},e^{u_{n-1}})-R_1(e^{u_{n+1}},e^{u_n})^{1/2}R_1(e^{u_{n-1}},e^{u_n})^{1/2}}{e^{u_{n+1}}-e^{u_{n-1}}},\\ R(u,v,w)=v^2\pm a(u+w)+uw,\quad R_1(v,u)=R(v,u,w).}
Уравнения с разными знаками связаны преобразованием $\tilde{u}_{n}=u_{n}+In\pi,$ которая не изменяет уравнение \eqref{list4}.

2.  При $b_1=b_2=0$ допускается еще одно дополнительное уравнение:
\mmeq{{u_{n+1,x}=\pm u_{n,x}+a_1e^{(\pm u_{n}+u_{n+1})/2}+a_2 e^{-(\pm u_{n}+u_{n+1})/2},\label{hyp_4_4}}}
\eq{u_{n,\tau}=\frac{e^{u_{n+1}/2}-e^{u_{n-1}/2}}{e^{u_{n+1}/2}+e^{u_{n-1}/2}}}
Уравнения с разными знаками связаны заменой $\tilde u_n=(-1)^nu_n,$ $a_1\leftrightarrow a_2$.

\itemaa Для уравнения \eqref{list5} полудискретное гиперболическое уравнение имеет вид:
\mmeq{{\arcsinh u_{n+1,x}-a\  \arcsinh u_{n,x}=g(u_{n+1}+bu_{n}), \label{hyp_2.5}}}
где параметры $a,b$ и функция $g(x)$ определяются из условий:
\eqa{&a=\pm1,\  b=\pm1,\ b_1(a+1)=0,\\ g(x)&=\ln(ab)+ \ln\frac{(y(x)+b_1+c)^2+2c(b_2-b_1)}{(y(x)+b_1-c)^2+2c(b_2+b_1)}, \\
y^{\prime}=&\frac{((y+b_1)^2+c(c+2b_2)-2cy)((y+b_1)^2+c(c+2b_2)+2cy)}{8c^{3/2}y}.}
Высшая симметрия в дискретном направлении имеет вид:
\eq{u_{n,\tau}=h_1+\frac{h_2(1-b)-2ah_1(1+b)}{2y(u_{n+1}+bu_n)y^b(u_n+bu_{n-1})-a(b+1)-a(1-b)(b_1^2+4b_2c+c^2)}\label{dis_5}.}
Она в зависимости от $a,b$ имеет представления:

1)При $a=1, b=1$ имеем $b_1=0$ и получаем уравнение (V9) так как уравнение на $y(x)$ имеет только четные степени неизвестной функции:
\eq{u_{n,\tau}=h_1\frac{y(u_{n+1}+u_{n})-y(u_{n}+u_{n-1})}{y(u_{n+1}+u_{n})+y(u_{n}+u_{n-1})}}

2)При $a=-1,b=1$ получаем уравнение (V10):
\eq{u_{n,\tau}=h_1\frac{y(u_{n+1}+u_{n})+y(u_{n}+u_{n-1})}{y(u_{n+1}+u_{n})-y(u_{n}+u_{n-1})}}

3)При $a=\pm 1,b=-1$  и получаем уравнение (V11):
\eqa{u_{n,\tau}=&h_1+\frac{ah_2}{2(b_1^2+2b_2c+c^2)}\frac{(y_1(u_{n+1}-u_n)+1)(y_1(u_n-u_{n-1})+1)}{y_1(u_{n+1}-u_{n})+y_1(u_{n}-u_{n-1})},\\
&y=\frac{y_1-1}{y_1+1}\sqrt{-a(b_1^2+2b_2c+c^2)},\\ y_1^\prime=&(a+1)\frac{c(y_1^2+1)^2+8b_2y_1^2}{8c^{1/2}(1-y_1^2)}\\+&(a-1)\frac{b_1^2(y_1^4+1)+b_1(y_1^4-1)\sqrt{b_1^2+2b_2c+c^2}+cb_2(y_1^2+1)^2+2c^2y_1^2}{4c^{3/2}(1-y_1^2)}.}

\itemaa Для уравнения Кричивера-Новикова \eqref{list7} получаем полудискретное уравнение
\[\refstepcounter{meq} \begin{array}{c}
u_{n+1,x}u_{n,x}=a_1u_{n+1}^2u_{n}^2+a_2u_{n+1}u_{n}(u_{n+1}+u_{n})\\+a_3(u_{n+1}^2+u_{n}^2)+a_5u_{n+1}u_{n}\\+a_6(u_{n+1}+u_{n})+a_9,\end{array}\eqno{(\hbox{S}\themeq)}\label{dhyp1}\]
коэффициенты уравнений связаны формулами
\eqa{b_4&=-6a_1a_3+3a_2^2/2,\quad b_3=-6a_1a_6-6a_2a_3+3a_2a_5,\\ b_2&=-6a_2a_9-3a_2a_6-6a_3^2+3a_5^2/2,\\ b_1&=-6a_2a_9-6a_3a_6+3a_5a_6,\quad b_0=-6a_3a_9+3a_6^2/2.}
Уравнение \seqref{dhyp1} найдено в работе \cite{a98}. Его высшая симметрия в  $n$ направлении является известной дискретизацией уравнения Кри\-че\-ве\-ра-Новикова \cite{y06}:
\eqa{u_{n,\tau}=&\frac{R(u_{n+1},u_{n},u_{n-1})}{u_{n+1}-u_{n-1}},\quad
R(u,v,w)=2uw(a_1v^2+a_2v+a_3)\\&+(u+w)(a_2v^2+a_5v+a_6)+2a_3v^2+2a_6v+2a_9.}

\itemaa Для уравнения \eqref{list6} есть разные подслучаи в зависимости от функции $Q(x)$. При $Q(x)=0$ уравнение \eqref{list6} совпадает с уравнением \eqref{list8}, которое описано ниже. При $\Q(x)\neq 0$ все полудискретные уравнения  записываются в виде:
\mmeq{\arcsinh\frac{u_{n+1,x}}{\sqrt{Q(u_{n+1})}}-a\arcsinh\frac{u_{n,x}}{\sqrt{Q(u_{n})}}=\arccosh\frac{aA(u_{n+1},u_{n})}{\sqrt{Q(u_{n+1})}\sqrt{Q(u_{n})}},\ a=\pm1.\label{d7}}
 Почти все дискретные симметрии имеют представление:
\eqa{u_{n,\tau}=\frac{R(u_{n-1},u_{n},u_{n+1})+\nu R(u_{n-1},u_{n},u_{n-1})^{1/2}R(u_{n+1},u_{n},u_{n+1})^{1/2}}{u_{n+1}-u_{n-1}}}
Для таких представлений функции $A$ и $R$ являются полиномиальными. Выпишем их для пяти канонических форм полинома $Q(x)$:

\itemnum При $Q(x)=x(x-1)(x-k)$ есть 4 разных ответа:
    \eqa{A(u,v)=b(u-v)^2-uv/2(u+v+2k+2)-k/2(u+v),\\ \nu=a,\\
	R(v,u,w)=(u+2b)^2vw+(-2bu^2-(4b^2-4bk-4b+k)u\\-2kb)(v+w)+(2bu-k)^2;}
\eqa{A(u,v)=b(k(u+v-1)-uv)^2-k/2(u^2+v^2-u-v)\\+uv/2(u+v-2),\\ \nu=-a,\\
	R(v,u,w)=(2bu+2bk-1)^2vw+(-2b(2bk-1)u^2\\+(4b^2k^2+4b^2k-4b-1)u-2bk(2bk-1))(v+w)\\+(2bku-2bk-u)^2;}
\eqa{A(u,v)=b(uv-k)^2+uv/2(u+v-2)+k/2(u+v-2uv),\\ \nu=-a,\\
	R(v,u,w)=(2bu+1)^2vw+(2bu^2-(4b^2k+4bk+4b+1)u\\+2bk)(v+w)+(2bk+u)^2;}
	\eqa{A(u,v)=b(uv-u-v+k)^2+u^2/2(v-1)+v^2/2(u-1)\\+k/2(u+v-2uv),\\ \nu=-a,\\
	R(v,u,w)=(2bu-2b+1)^2vw+(-2b(2b-1)u^2\\+(4b^2k+4b^2-4b-1)u-2bk(2b-1))(v+w)\\+(2bk-2bu+u)^2;}
 
\itemnum  При $Q(x)=x(x-1)$ есть 3 разных ответа. Два из них $S$-интегрируемые
\eqa{A(u,v)=(u-v)^2b-uv+1,\quad \nu=a,\\R(v,u,w)=b^2vw-bu(b+1)(v+w)+b^2u^2+2b+1;}
\eqa{A(u,v)=(u+v)^2b-uv-1,\quad \nu=-a\\R(v,u,w)=b^2vw+bu(b-1)(v+w)+b^2u^2-2b+1.}
И одно Дарбу интегрируемое уравнение при 
\eqa{A(u,v)=b+uv,}с интегралами
\eqa{&W_1=\frac{(z_1+\sqrt{z_1^2-(u_n^2-1)(b^2-1)})(z_2+\sqrt{z_2^2-(u_n^2-1)(b^2-1)})^a}{(u_n^2-1)^{(a+1)/2}},\\&z_1=bu_n+u_{n-1},\ z_2=bu_n+u_{n+1},\\&W_2=a^n\frac{u_{n,xx}+u_{n}}{\sqrt{u_{n,x}^2+u_{n}^2-1}}}

\itemnum  При $Q(x)=x^2$ есть 2 разных ответа:
\eqa{A(u,v)=b(cuv+1)^2+uv,\quad\nu=-a\\ R(v,u,w)=bc^2u^2vw+u(bc+1)(v+w)+b;}
\eqa{&A(u,v)=b(cv+u)^2-uv,\\ &R(v,u,w)=bc(vw+u^2)+u(bc-1)(v+w);\\&u_\tau=\frac{R(u_{n-1}/c,u_{n},cu_{n+1})-a R(u_{n-1}/c,u_{n},u_{n-1}/c)^{1/2}R(cu_{n+1},u_{n},cu_{n+1})^{1/2}}{c^2u_{n+1}-u_{n-1}}}
С помощью неавтономной замены  $u_n=c^{-n}v_n$ можно добиться значения $c=1$.

\itemnum  При $Q(x)=x$ есть 2 разных ответа:
\eqa{A(u,v)=b(u-v)^2/2-u/2-v/2,\\ R(v,u,w)=b^2vw-b(bu+1)(v+w)+(bu-1)^2;}
и интегрируемое по Дарбу уравнение при 
\eqa{A(u,v)=b/2+u/2+v/2
.} Его интегралы имеют вид \eqa{W_1=&\frac{\left(z_1+\sqrt{z_1^2+8bu_n}\right)\left(z_2+\sqrt{z_2^2+8bu_n}\right)^a}{u_n^{(a+1)/2}},\\ &z_1=2b-u_n+u_{n-1},\ z_2=2b-u_n+u_{n+1}\\ W_2=&a^n\frac{2u_{n,xx}+1}{\sqrt{u_{n,x}^2+u_{n}}}}

\itemnum  При $Q(x)=1$ есть 2 разных ответа:
\eqa{A(u,v)=b(u+v+c)^2+1,\\ R(v,u,w)=b(v+u+c)(w+u+c)+2;}
\eqa{&A(u,v)=b(u-v-c)^2-1,\\ &u_\tau=\frac{2+b(c+u_{n-1}-u_n)(c+u_n-u_{n+1})}{u_{n+1}-u_{n-1}-2c}\\&\quad+\frac{a(2-b(c+u_{n-1}-u_n)^2)^{1/2}(2-b(c+u_n-u_{n+1}))^{1/2}}{u_{n+1}-u_{n-1}-2c}.}С помощью неавтономной замены  $u_n=v_n+cn$ можно добиться значения $c=0$.

\itemaa Уравнения \eqref{list8} можно рассматривать как частный случай уравнения \eqref{list6}, поэтому ответ содержится в формуле \eqref{dhyp1}:
\mmeq{u_{n+1,x}u_{n,x}=(au_{n+1}u_n+bu_{n+1}+cu_n+d)^2,}
\eqa{u_{n,\tau}=\frac{h_1(au_{n+1}u_n+bu_{n+1}+cu_n+d)(au_{n}u_{n-1}+bu_{n}+cu_{n-1}+d)}{a(b-c)u_{n+1}u_{n-1}+(b^2-ad)u_{n+1}-(c^2-ad)u_{n-1}+d(b-c)}\\+h_2(au_{n}^2+(b+c)u_{n}+d)}
Эти уравнения в переменной $$v_n=\frac{2au_n-A+b+c}{2au_n+A+b+c}\left(\frac{A+b-c}{A-b+c}\right)^n,\quad A=\sqrt{(b+c)^2-4ad}$$имеют наиболее простой вид:\eqa{v_{n+1,x}v_{n,x}=(v_{n+1}+v_{n})^2(bc-ad),\\ v_{n,\tau}=h_1\frac{(v_{n+1}+v_{n})(v_{n}+v_{n-1})}{v_{n+1}-v_{n-1}}+h_2Av_n}

А также Дарбу интегрируремые уравнения:
\mmeq{u_{n+1,x}=u_{n,x}A(u_{n+1,0},u_{n,0}),\quad A_y(z,y)+A(z,y)A_z(z,y)=0}
\eqa{W_1=A(u_{n+1},u_{n}),\quad \frac{d}{dx}W_1=0,\\
W_2=u_{n,xx}/u_{n,x},\quad TW_2=W_2.}
\eqa{u_{n+1,x}=\frac{u_{n,x}}{(A(u_{n+1,0},u_{n,0})u_{n,0}+g(A(u_{n+1,0},u_{n,0})))^2},\\ A_y(z,y)+\frac{A_z(z,y)}{(A(z,y)z+g(A(z,y)))^2}=0,}
\eqa{W_1=A(u_{n+1},u_{n}),\quad \frac{d}{dx}W_1=0,\\
W_2=2\frac{u_{n,xxx}}{u_{n,x}}-3\frac{u_{n,xx}^2}{u_{n,x}^2},\quad TW_2=W_2.}
Такие уравнения были найдены в работе \cite{s12}.

\itemaa Для уравнения \eqref{list9} ответы существуют лишь при $b_1=0$ и имеют вид
\mmeq{\sqrt{u_{n+1,x}}\pm\sqrt{u_{n,x}}=\pm b_2\sqrt{(u_{n}-u_{n+1}+a)^2-b}\label{hyp9}}
\eqa{u_{n,\tau}=h_1\frac{R(u_{n-1},u_{n},u_{n+1})-\sqrt{R_1(u_{n-1},u_{n})}\sqrt{R_1(u_{n+1},u_{n})}}{u_{n+1}-u_{n-1}-2a}+h_2,\\ R(v,u,w)=(u-v-a)(u-w+a)-b,\quad R_1(v,u)=R(v,u,v).}
Константу $a$ можно убрать неавтономным преобразованием $\tilde u_n=u_n+an$

\section{Обсуждение результатов.}

Ряд примеров полученных в предыдущем параграфе можно представить в одной из следующих форм:
\eqs{\frac{d}{dx}\psi(u_{n+1},u_{n})=\phi(u_{n+1},u_{n})\label{z_up}} или
\eqs{\Psi(u_{n+1,x},u_{n+1})=\Phi(u_{n,x},u_{n}).\label{z_down}} Например, в виде \eqref{z_up} можно записать уравнения (\sref{hkdv}, \sref{hyp_2}, \sref{hyp_2.1}, \sref{hyp_3}, \sref{hyp_4_1}, \sref{hyp_4_3}, \sref{hyp_4_4}), а в виде \eqref{z_down} уравнения (\sref{hyp_2},\sref{hyp_2.1}) при $a=0$, (\sref{hyp_4_3},\sref{hyp_4_4}) при $a=\pm \sqrt{c_1c_2}$ и $a=\pm1$ соответственно, уравнения вида \seqref{d7} также можно представить в таком виде при специальном выборе параметров (когда функция $R(v,u,v)$ является полным квадратом). Некоторые из уравнний допускают оба представления.

Для уравнений вида \eqref{z_up} и \eqref{z_down} можно ввести замены переменных обратимые на решениях. Так, для уравнения \eqref{z_up} можно обозначить $$v_{n}=\psi(u_{n+1},u_{n}),$$ тогда в силу \eqref{z_up} имеем $$v_{n,x}=\phi(u_{n+1},u_{n}).$$ В случаях, когда существует обратная замена $$u_{n}=\tilde\psi(v_{n,x},v_{n}),\ u_{n+1}=\tilde\phi(v_{n,x},v_{n}),$$ на новую функцию $v_n$ получаем полудискретное уравнение вида \eqref{z_down}: $$\tilde\psi(v_{n+1,x},v_{n+1})= \tilde\phi(v_{n,x},v_{n}).$$
Видно, что аналогичным образов можно строить замены от уравнений вида  \eqref{z_down} к уравнениям вида \eqref{z_up}. Замены такого вида рассматривались ранее в \cite{y90,y93,s14}. При таких заменах высшие симметрии также пересчитываются \cite{y90}. Поэтому часть из перечисленных уравнений содержится в работе \cite{y90}, в частности, уравнения вида \seqref{d7} в случаях, когда функция $R(v,u,v)$ является полным квадратом. Остальные уравнения кроме \seqref{dhyp1} скорее всего являются новыми. В частности, уравнения \seqref{hyp_2.5}, \seqref{d7}, \seqref{hyp9} в ситуациях общего положения.

Автор статьи выражает благодарность анонимному рецензенту доктору за критические замечания и ценные советы.


\bigskip
\begin{thebibliography}{999}
\bibitem{vs93}А.П. Веселов, А.Б. Шабат, \emph{Одевающая цепочка и спектральная теория оператора Шрёдингера}, Функц. анализ и его прил., {\bf 27}:2, 1–21 (1993). 

\bibitem{y06} R. Yamilov,  \emph{Symmetries as integrability criteria for differential difference equations}. Journal of Physics A: Mathematical and General, {\bf 39}(45), R541 (2006).

\bibitem{y83}Р.И. Ямилов, \emph{О классификации дискретных эволюционных уравнений}, Успехи мат. наук {\bf 38}:6, 155-156  (1983).

\bibitem{y90} Р.И. Ямилов, {\emph Обратимые замены переменных, порожденные преобразованиями Беклунда}, Теор. и мат. физ. {\bf 85}:3, 368-375 (1990).

\bibitem{gh21}R.N.Garifullin and I.T.Habibullin \emph{Generalized symmetries and integrability conditions for hyperbolic type semi-discrete equations}, Journal of Physics A: Mathematical and Theoretical, {\bf 54}:20, 205201, 19 pp (2021).


\bibitem{ms12} А.Г. Мешков, В.В. Соколов, \emph{Интегрируемые эволюционные уравнения с постоянной сепарантой}, Уфимск. матем. журн., {\bf 4}:3,  104–154 (2012).  [Engl. trans.: Ufa Math. Journal {\bf 4}:3. 104--152 (2012).]

\bibitem{ss82}С. И. Свинолупов, В. В. Соколов, \emph{Об эволюционных уравнениях с нетривиальными законами сохранения}, Функц. анализ и его прил., {\bf 16}:4 (1982), 86–87; Funct. Anal. Appl., 16:4 (1982), 317--319

\bibitem{gy12}R.N. Garifullin and R.I. Yamilov, \emph{Generalized symmetry classification of discrete
equations of a class depending on twelve parameters}, J. Phys. A: Math. Theor. {\bf 45} (2012) 345205 (23pp).

\bibitem{gmy14}Р. Н. Гарифуллин, А. В. Михайлов, Р. И. Ямилов, \emph{Дискретное уравнение на квадратной решетке с нестандартной структурой высших симметрий}, ТМФ, {\bf 180}:1 (2014), 17–34.

\bibitem{gy19}Р. Н. Гарифуллин, Р. И. Ямилов, \emph{Необычная серия автономных дискретных интегрируемых уравнений на квадратной решетке}, ТМФ, {\bf 200}:1 (2019), 50--71.

\bibitem{a98}V. E. Adler, \emph{B\"acklund transformation for the Krichever-Novikov equation}, International Mathematics Research Notices, {\bf 1998}:1, 1-4 (1998).

\bibitem{s12}С. Я. Старцев, \emph{Интегрируемые по Дарбу дифференциально-разностные уравнения, допускающие интеграл первого порядка}, Уфимск. матем. журн., {\bf 4}:3 (2012), 161-176


\bibitem{y93} R.I. Yamilov, {\emph On the construction of Miura type
transformations by others of this kind}, Phys. Lett. A {\bf 173} (1993) 53--57.


\bibitem{s14} S.Ya. Startsev, \emph{Non-Point Invertible Transformations and Integrability of Partial Difference Equations}, SIGMA, {\bf 10} (2014), 066, 13 pages.

\end{thebibliography}
\end{document}